\documentclass[12pt]{article}

\usepackage[dvips]{graphics}
\usepackage{booktabs,lscape,amssymb,amsmath}

\newcommand{\pal}{\partial}

\newcommand{\be}{\begin{equation}}
\newcommand{\ee}{\end{equation}}
\newcommand{\bea}{\begin{eqnarray}}
\newcommand{\eea}{\end{eqnarray}}
\newcommand{\bi}{\begin{itemize}}
\newcommand{\ei}{\end{itemize}}
\newcommand{\bay}{\begin{array}}
\newcommand{\eay}{\end{array}}

\newcommand{\non}{\nonumber}
\newcommand{\Dw}{D_{\rm w}}

\newcommand{\Dov}{D_{\rm N}}
\newcommand{\pbf}{{\bf p}}
\newcommand{\hq}{\hat{q}}
\newcommand{\tq}{\tilde{q}}

\title{
\begin{flushright}
{\normalsize HU-EP-07/61}\\
{\normalsize SFB/CPP-07-84}\\
\end{flushright}
Perturbative analysis of the Neuberger-Dirac
operator in the Schr\"odinger functional
}

\author{Shinji Takeda\\
Humboldt Universit\"at zu Berlin,\\
Newtonstr.~15, 12489 Berlin, Germany.}

\begin{document}

\maketitle
\abstract{
We investigate the spectrum of the free Neuberger-Dirac operator $\Dov$
on the Schr\"odinger functional (SF).
We check that the lowest few eigen-values of
the Hermitian operator $\Dov^{\dag}\Dov$ in unit of $L^{-2}$
converge to the continuum limit properly.
We also perform a one-loop calculation of the
SF coupling, and then check the universality and investigate
lattice artifacts of the step scaling function.
It turns out that the lattice artifacts for the Neuberger-Dirac operator
are comparable in those of the clover action.
}

\section{Introduction}
Chiral symmetry plays an important role
in the understanding of the strong interaction.
A solution to a realization of the exact chiral symmetry
on the lattice is proposed by Neuberger \cite{Neuberger:1997fp}
as the Neuberger-Dirac operator (overlap).
Recently, dynamical overlap lattice QCD simulations
started in Ref.~\cite{Fodor:2003bh} and
see \cite{Schaefer:2006bk} for an overview of recent progress.
Nowadays, thanks to developments of algorithms
and powerful current computers,
large scale simulations are feasible as shown
by the JLQCD collaboration in Ref.~\cite{Fukaya:2007fb}.
In that course,
even after removing systematic errors
\footnote{Of course, quenching is also one of the main sources
of systematic errors.},
finite size effects,
cutoff effects and
an ambiguity of chiral extrapolation,
non-perturbative renormalization becomes an
essential element
for accurate quantitative predictions.

One elegant solution to this issue is
the Schr\"odinger functional (SF) scheme \cite{Luscher:1992an}
which is an intermediate scheme
connecting the perturbative and hadronic regime.
This method was shown to be useful to study
the non-perturbative evolution
over a wide range for various quantities,
the coupling constant \cite{Luscher:1993gh},
the quark masses \cite{Capitani:1998mq},
the structure function \cite{Guagnelli:1999gu,Guagnelli:2003hw}, and
the weak matrix elements \cite{Guagnelli:2005zc}.
By making use of the scaling technique,
one can complete a perturbative matching safely
at relatively high energy, and then
the renormalization group invariant quantities, like
the lambda parameter \cite{Dellamorte:2005},
the masses for the light quark \cite{DellaMorte:2005kg}
and the heavy quark by HQET \cite{DellaMorte:2006cb},
are determined without worrying about the systematic error of
truncation of the perturbative expansion.
Although so far the method was mainly used
for Wilson type fermions \cite{Sint:1993un},
there are several attempts for other fermion formulations,
like staggered fermions
\cite{Miyazaki:1994nu,Heller:1997vh,PerezRubio:2007qz}
and domain wall fermions \cite{Taniguchi:2006qw}.
Recently, formalisms for the Neuberger-Dirac operator
on the SF have been proposed by Taniguchi
\cite{Taniguchi:2004gf} and L\"uscher \cite{Luscher:2006df}.
The former employs the orbifolding technique
and the latter is based on universality considerations.
In Ref.~\cite{Leder:2007aq,Leder:2007nj},
the latter formulation is examined
in the framework of the Gross-Neveu model.
Here in this paper, we implement the formulation in QCD
and study the spectrum of the free operator and calculate
the SF coupling to one-loop order.
Furthermore we investigate lattice artifacts
of the step scaling function by comparing with
the clover fermion action.

The rest of the paper is organized as follows.
In Section \ref{sec:def}, we summarize the definition
of the Neuberger-Dirac operator on the SF
which is given by L\"uscher, and
give some practical details about building the operator.
Then we show the spectrum of
the free Neuberger-Dirac operator in Section \ref{sec:SpecND}.
In Section \ref{sec:SFC},
we present results for the fermion part
of the SF coupling to one-loop order
by making use of the Neuberger-Dirac operator.
Furthermore we investigate lattice artifacts of the step scaling
function in Section \ref{sec:SSF}.
Finally, we give some concluding remarks and outlook
in Section \ref{sec:conclusion}.
In Appendices, we give an explicit form
of the Wilson-Dirac operator on the SF in a
time-momentum space for later use (\ref{sec:Dw}),
and a discussion about a determination
of a boundary coefficient at tree level (\ref{sec:ctree}),
and we summarize some tables of numerical results (\ref{sec:tables}).

\section{Neuberger-Dirac operator on the SF}

\subsection{Definition}
\label{sec:def}
A massless Neuberger-Dirac operator
on the SF (with size $T\times L^3$) \cite{Luscher:2006df} is defined by
\bea
\Dov
&=&
\frac{1}{\bar{a}}
\{ 1 - \frac{1}{2} ( U + \tilde{U}) \},
\label{eqn:Dov}
\\
U
&=&
A X^{-1/2},
X=A^{\dag}A+c a P,
\label{eqn:U}
\\
\tilde{U}
&=&
\gamma_5 U^{\dag} \gamma_5,
\label{eqn:tildeU}
\eea
with $\bar a=a/(1+s)$.
The operator follows the modified Ginsberg-Wilson (GW) relation
\be
\gamma_5 \Dov + \Dov \gamma_5
=
\bar{a} \Dov \gamma_5 \Dov + \Delta_{\rm B},
\ee
where $\Delta_{\rm B}$ is the exponentially local operator.
The $A$ in the kernel operator $X$
of the inverse square root is given by
\be
A=1+s-a\Dw,
\ee
where $\Dw$ is the massless Wilson fermion on the SF \cite{Sint:1993un}.
The tunable parameter $s$ is taken in a range
$-0.6\le s \le 0.6$ in the following.
The boundary coefficient $c$ represents the strength
of the boundary operator $P$ which is given by
\be
P
=
\frac{1}{a}
\delta_{\bf x,y}
\delta_{x_0,y_0}
\{
 \delta_{x_0,a}P_-
+\delta_{x_0,T-a}P_+
\}.
\label{eqn:P}
\ee
In Ref.~\cite{Luscher:2006df},
it is shown that the kernel operator
$X=A^{\dag}A+ c a P$ is bounded from below by
the spectral gap of $A^{\dag}A$ on the infinite lattice
if $c\ge 1$ holds,
and furthermore it is mentioned that
\be
c=1+s,
\label{eqn:ctreeL}
\ee
is the nearly optimal choice
in order to achieve tree level $O(a)$ improvement.
We investigate this point in some detail in
Appendix \ref{sec:ctree}, and we conclude that,
for the precision of our calculation here (and maybe for future simulations),
this formula is accurate enough.
Therefore in the following calculations, apart from the appendix,
we alway set $c$ to the value in eq.(\ref{eqn:ctreeL}).
In this paper, we restrict ourself to the massless case
of the Neuberger-Dirac operator.

To carry out perturbation calculations,
it usually is convenient to move to momentum space.
In the SF setup, however, the Fourier transformation
can be done only in the spatial directions,
due to a lack of translation invariance for the time direction.
Therefore, we work in a time-momentum space,
\be
\psi(x_0,{\bf p})
=
a^3\sum_{\bf x} e^{-i{\bf px}} \psi(x).
\ee
The explicit expression of the free $\Dw$ in time-momentum space
with the Abelian background gauge field
in the SU($3$) group \cite{Luscher:1993gh} is shown
in Appendix \ref{sec:Dw}.
Here we show a matrix expression of $A$ in time-momentum space
for a fixed spatial momentum $\pbf$ and a color $b$($=1,2,3$),
\be
A^b(\pbf)
=
\left[
\begin{array}{ccccccc}
g^b(\pbf;a)&  P_-       &  0         &\cdots&\cdots&  0        &  0        \\
P_+        &g^b(\pbf;2a)&  P_-       &0     &\cdots&  \cdots   &  0   \\
0          &  P_+       &g^b(\pbf;3a)&P_-   &0     &  \cdots   &  \vdots   \\
\vdots     &  0         &  P_+       &\ddots&\ddots&  \ddots   &  \vdots   \\
\vdots     &  \ddots    &  \ddots    &\ddots&\ddots&  P_-     &  0        \\
0          &  \cdots    &  \cdots    &0     &P_+   &g^b(\pbf;T-2a) & P_-  \\
0          &  0         &  \cdots    &\cdots&0     & P_+      &g^b(\pbf;T-a)\\
\end{array}
\right],
\label{eqn:matrixA}
\ee
which is a totally $4(T/a-1)\times 4(T/a-1)$ matrix.
Block elements $P_{\pm}=(1\pm \gamma_0)/2$ and $g^b(\pbf,x_0)$
have Dirac spinor structure, and the latter is given by
\bea
g^b(\pbf;x_0)
&=&
s
-\frac{1}{2}\sum_{k=1}^{3} \hq_k^b(x_0)^2
-i\sum_{k=1}^{3} \tq_k^b(x_0)\gamma_k,
\eea
where $\tilde q$ and $\hat q$ are function of the spatial
momentum $\pbf$, the time $x_0$ and $\theta$, and they are
defined in Appendix \ref{sec:Dw}.
The dependence of $b$ is caused by the color diagonal background field.
The angle $\theta$ comes from the generalized
periodic boundary condition for the spatial directions,
\be
\psi(x+L\hat{k})=e^{ i\theta}\psi(x),
\hspace{8mm}
\bar\psi(x+L\hat{k})=e^{-i\theta}\bar\psi(x),
\ee
for $k=1,2,3$.
In time-momentum space,
the boundary operator in eq.(\ref{eqn:P}) is represented as
\be
P_{\rm tm}
=
\frac{1}{a}
\delta_{x_0,y_0}
\{
 \delta_{x_0,a}P_-
+\delta_{x_0,T-a}P_+
\},
\ee
and it's matrix expression is given by
\be
P_{\rm tm}
=
\left[
\begin{array}{ccccccc}
P_-   & 0     & \cdots&\cdots&\cdots&0     \\
0     & 0     & \ddots&\ddots&\ddots&\ddots\\
\vdots& \ddots& \ddots&\ddots&0     &0     \\
0     & \cdots& \cdots&\cdots&0     &P_+   \\
\end{array}
\right].
\label{eqn:boundaryP}
\ee

As it is clear from the explicit form, even in the free case
we cannot have an analytic form of the
Neuberger-Dirac operator on the SF,
due to the presence of the background field.
Therefore we have to rely on an
approximation to the inverse square root, even for
perturbative calculations.
We will return to this issue how
to build the operator in Section \ref{subsec:build}.

\subsection{Distribution of $\epsilon$}
\label{sec:Distributionepsilon}

When one approximates the $X^{-1/2}$ by a polynomial of $X$
following Ref.~\cite{Hernandez:1998et},
one needs information about the lower $u$ and upper $v$ bound of $X$.
When the ratio $u/v$ is not too small,
one can obtain its approximation with lower degree.
Usually, $u$ and $v$ are set to the values of minimal and
maximal eigen-value (or norm)
of the $X$, therefore it is important to know the spectrum of $X$.
Here we discuss the spectrum of $X$ at tree level
in the presence of the background field.

At fixed momentum ${\bf p}$ and color $b$,
we evaluate the minimal eigen-value and the norm
of the kernel operator
\be
X^b(\pbf)=(A^b(\pbf))^{\dag}A^b(\pbf)+c a P_{\rm tm},
\ee
and set them as
the lower and upper range of the approximation
$[u^b(\pbf),v^b(\pbf)]$
\bea
u^b(\pbf)
&=&
\lambda_{\min}(X^b(\pbf)),
\\
v^b(\pbf)
&=&
||X^b(\pbf)||.
\eea
Essentially, $\epsilon^b(\pbf)=u^b(\pbf)/v^b(\pbf)$
controls the cost of the computation,
since for a given precision
it determines the degree ($N$) of the approximation polynomial.

\begin{figure}[!ht]
 \begin{center}
  \begin{tabular}{cc}
   \scalebox{1.0}{\includegraphics{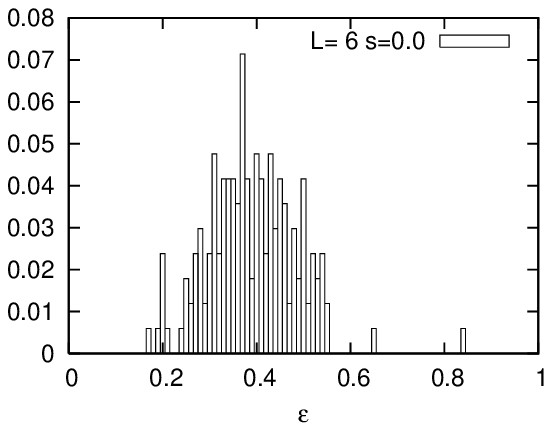}}&
   \scalebox{1.0}{\includegraphics{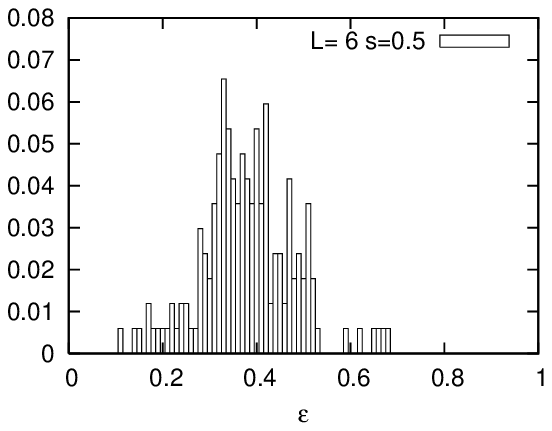}}\\
   \scalebox{1.0}{\includegraphics{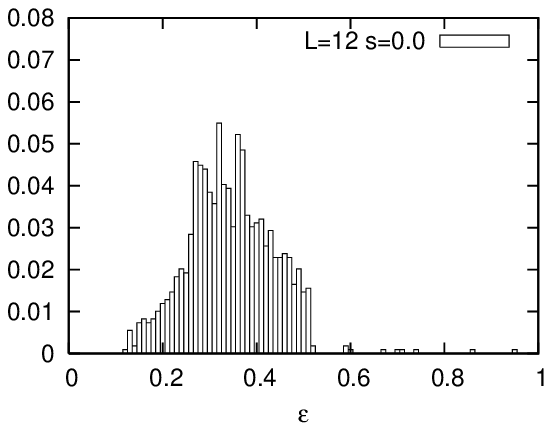}}&
   \scalebox{1.0}{\includegraphics{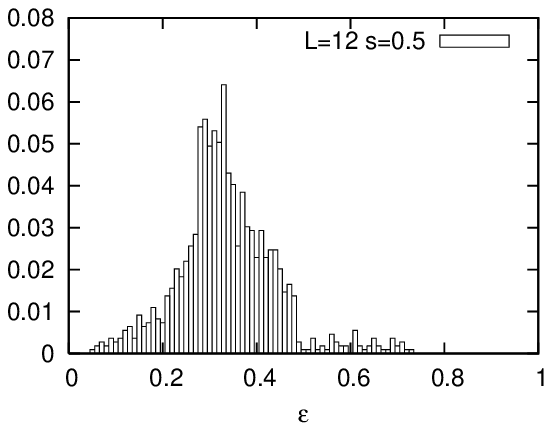}}\\
   \scalebox{1.0}{\includegraphics{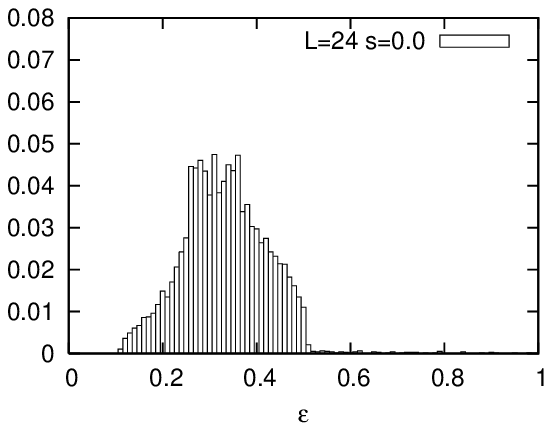}}&
   \scalebox{1.0}{\includegraphics{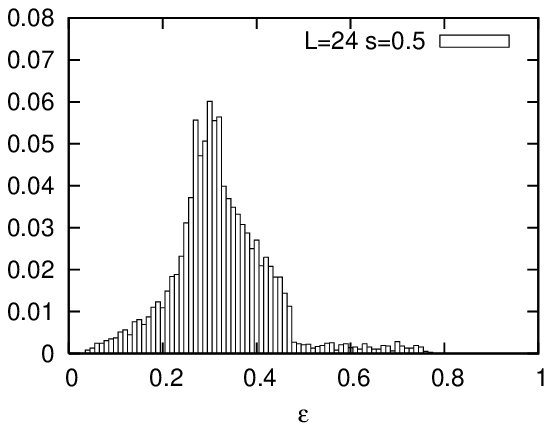}}\\
  \end{tabular}
 \end{center}
 \caption{We show the distribution of
$\epsilon^b(\pbf)=u^b(\pbf)/v^b(\pbf)$
over the all spatial momenta $\pbf$ and the color $b=1,2,3$
for the case of $\theta=0$ with the non-zero background field.
We show here for lattice sizes $L/a=T/a=6,12,24$ (from top to bottom) and
$s=0.0$ (Left panels) and $s=0.5$(Right panels).
Due to the permutation symmetry for the spatial momentum,
we only have to investigate a momentum configurations $p_3\le p_2 \le p_1$.
We did not correct the weight in the histogram,
therefore the distribution shown here
is not fully correct one.
However the distribution is practically meaningful,
since we do not do a computation on a full configurations with size $(L/a)^3$,
but only on the configurations $p_3\le p_2 \le p_1$
(about $(L/a)^3/6$).
\label{fig:epsilon}
}
\end{figure}

Its distribution over $\pbf$ and color $b$
in the case of $\theta=0$ with the non-zero background field
are shown in Figure \ref{fig:epsilon} for
several lattice sizes $L/a=T/a=6,12,24$ and $s$ parameters $s=0.0,0.5$.
As you can see in the figure, $\epsilon$ is distributed
in a relatively large range $0.01\lesssim\epsilon\lesssim 1$,
therefore we concluded that it is better to calculate
coefficients of the polynomial expansion
for each $\pbf$ and $b$ for saving time.
The calculation of the coefficients
is much cheaper than summing up the polynomial expansion.
(You can use common lower and upper bound
$u_{\rm comm}=\min_{\pbf,b}\{\lambda_{\rm min}(X^b(\pbf))\}$ and
$v_{\rm comm}=\max_{\pbf,b}||X^b(\pbf)||$ for all momentum and color sector,
but this is clearly inefficient.)

\subsection{How to build the Neuberger-Dirac operator}
\label{subsec:build}

To build the inverse square root
in the Neuberger-Dirac operator,
we adopt the Chebyshev polynomial approximation,
\be
f(X)=X^{-1/2}
\sim
f_N(X)=\sum_{k=0}^{N}c_kT_k(Y),
\label{eqn:polynomialsum}
\ee
where $Y=[2X-(v+u)]/(v-u)$ for the lower $u$ and upper $v$
bound of the $X$,
and $T_k$ is the Chebyshev
polynomial of degree $k$.
Main tasks to obtain the operator are
two-fold: the first is a computation of the coefficients $c_k$
and the second is to sum up in eq.(\ref{eqn:polynomialsum}).
Concerning the latter part,
we use the Clenshaw summation scheme
to maintain precision.
For the former part,
we examine two methods to compute the $c_k$,
the Remez algorithm and the Chebyshev interpolation,
in order to check rounding off errors
occurring in the perturbative calculation in Section
\ref{sec:SFC}.

Following Ref.~\cite{Giusti:2002sm},
we implement the Remez algorithm to obtain minimax polynomials.
On the other hand, we follow the Numerical Recipes
for the Chebyshev interpolation.
In calculations of the spectrum of the Neuberger-Dirac operator
in Section \ref{sec:SpecND},
we only use the Chebyshev polynomial method.
While for computations
of the one-loop coefficient of the SF coupling in Section \ref{sec:SFC},
we employ both methods and show the results
to digits which are common in both methods.

When approximating the inverse square root of $X$,
we demand a precision
\be
\max_{u\le x \le v} \left|\frac{f(x)-f_N(x)}{f(x)}\right|
< 10^{-13},
\label{eqn:precision}
\ee
and we alway check consistency when it is built
\be
X\times (X^{-1/2})^2 = I,
\ee
and we observe that errors on the right-hand side
are less than $2\times 10^{-13}$.


\section{Spectrum of the free Neuberger-Dirac operator}
\label{sec:SpecND}

\subsection{Spectrum of $\Dov$}
\label{sec:SpecDov}
Since the $U$ in eq.(\ref{eqn:U}) is not a unitary matrix,
there is no guarantee that its spectrum is distributed
on a unit circle whose origin is $(1,0)$ as in the case of
the infinite volume.
However on the SF,
as it is shown in Ref.~\cite{Luscher:2006df}, since
\be
||U||=||\tilde U||\le 1,
\ee
one can see that
\be
|| \bar{a}\Dov - 1 ||
=
\frac{1}{2}|| U+\tilde U ||
\le 1.
\ee
Thus the spectrum of $\bar{a}\Dov$ is contained in a unit disk
which is enclosed by the unit circle.

\begin{figure}[!t]
 \begin{center}
  \begin{tabular}{cc}
  \scalebox{1.2}{\includegraphics{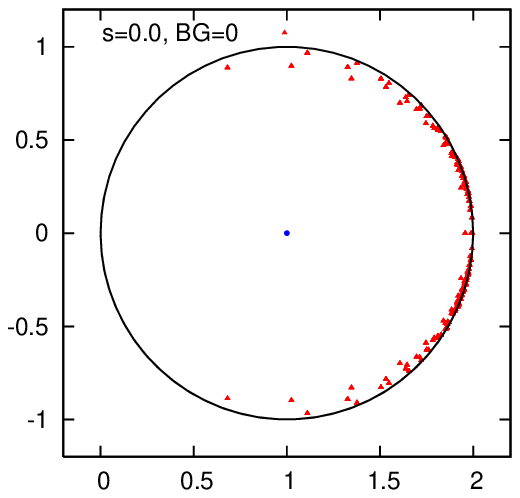}}&
  \hspace{-30mm}
  \scalebox{1.2}{\includegraphics{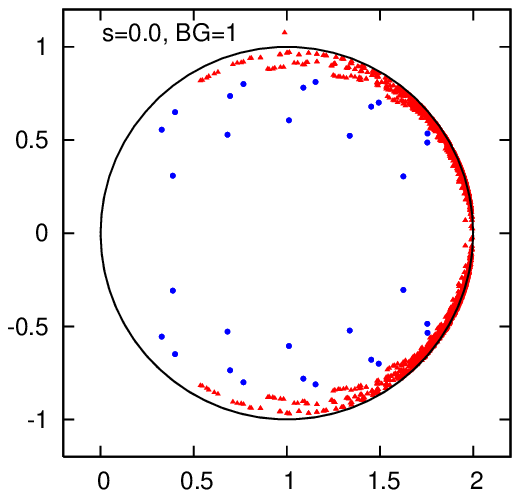}}\\
  \scalebox{1.2}{\includegraphics{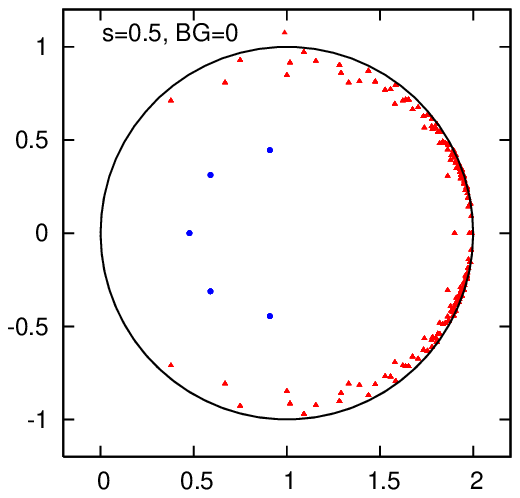}}&
  \hspace{-30mm}
  \scalebox{1.2}{\includegraphics{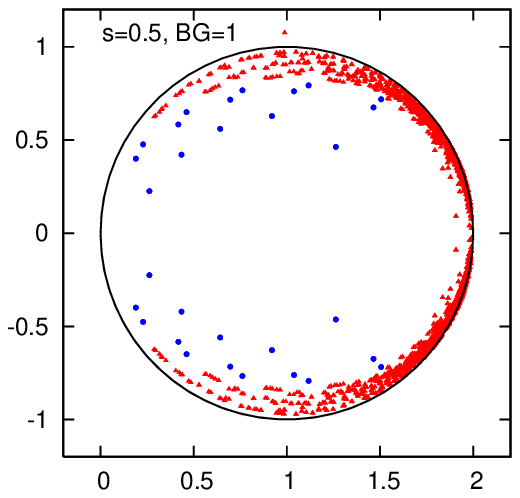}}\\
  \end{tabular}
 \end{center}
 \caption{Spectrum of $\bar{a}\Dov$ with parameters $\theta=0$
and for lattice size $L/a=T/a=6$.
The upper panels are for $s=0.0$,
while the lower ones are for $s=0.5$.
The value of BG means that $0$: zero background gauge field,
$1$: non-zero background gauge field (choice A).
All eigen-values are enclosed by a black circle whose
origin is $(1,0)$ and radius is $1$,
on which the spectrum of the GW relation operator lie.
The blue circle points which come from the
${\bf p}=0$ sector are located
far away from the circle, and they are positioned
around a center of the circle.
We observe that this sector is strongly
affected by boundary effects.
\label{fig:Dov}}
\end{figure}

We show the actual distribution of the spectrum in the free case
in Figure \ref{fig:Dov}
for $s=0.0$ and $s=0.5$.
In the figures, we only show
$\theta=0$ case, but $\theta=\pi/5$ is
also computed and has a similar tendency.
In the figures, results for
lattice size $L/a=T/a=6$ and with the zero (BG$=0$) or non-zero (BG$=1$)
background field are shown.
Most eigen-values are localized
near the unit circle, and a remnant
of the GW relation is observed.
Especially, note that ${\bf p=0}$ is most strongly affected by
the boundary.
When switching on the background gauge field (BG$=1$),
some degeneracies are lifted, and
you can see 'more' points than at zero background field case (BG$=0$).

\subsection{Spectrum of $\Dov^{\dag}\Dov$}
\begin{figure}[!ht]
 \begin{center}
  \begin{tabular}{cc}
  \scalebox{1.6}{\includegraphics{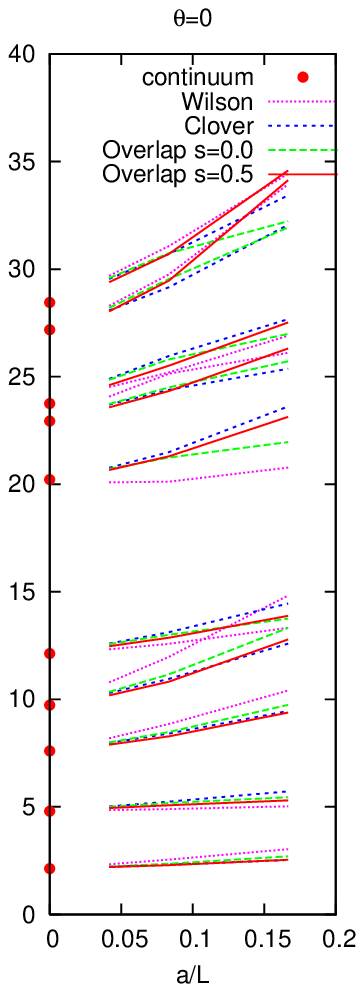}}&
  \scalebox{1.6}{\includegraphics{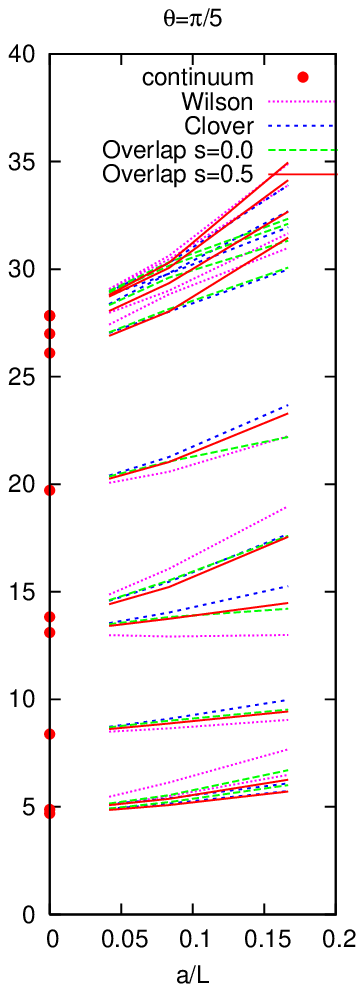}}
  \end{tabular}
 \end{center}
 \caption{The $a/L$ dependence of the lowest 10 eigen-values
of $L^2 D^{\dag}D$ for the various fermion actions in the presence of
the background field.
The left (Right) panel is for $\theta=0$ ($\theta=\pi/5$).
We compute the eigen-value of the Neuberger-Dirac (overlap) operator
for $s=0.0,0.1,0.2,0.3,0.4,0.5$, but
here we only show $s=0.0,0.5$, since
the others($s=0.1,0.2,0.3,0.4$)
just interpolate in between $s=0.0$ and $s=0.5$.
Those of the Wilson-Dirac action and the clover action
from \cite{Sint:1995ch} are shown for comparison.
The eigen-values converge to
the continuum ones (Red points at $a/L=0$) in Ref \cite{Sint:1995ch}.
Within the lowest 10 eigen-values,
no level crossing occurs in the Neuberger-Dirac operator unlike the case
of the Wilson-Dirac fermions.
\label{fig:gam5Dov2}}
\end{figure}

We also investigate the spectrum of the Hermitian operator
$L^2\Dov^{\dag}\Dov=(L/a)^2(1+s)^2 \bar{a}\Dov^{\dag}\bar{a}\Dov$,
because we can take the continuum limit
and compare the scaling behavior
with that of the Wilson-Dirac operator,
$L^2 \Dw^{\dag}\Dw=(L/a)^2 a D_{\rm w}^{\dag}a D_{\rm w}$
\cite{Sint:1995ch}.
The numerical results of the lowest
10 eigen-values of $L^2 \Dov^{\dag}\Dov$
with non-zero background field are summarized in Table \ref{tab:MIN10}
for $s=0.0,0.5$, $\theta=0,\pi/5$ and $L/a=T/a=6,12,24$.
The scaling to the continuum limit are plotted in Figure
\ref{fig:gam5Dov2} including those of the Wilson-Dirac and the clover
action ($c_{\rm sw}=1$) for comparison.
In the figure, the lower modes show a good scaling behavior,
while higher modes are strongly affected by lattice artifacts.

\section{SF coupling to one-loop order}
\label{sec:SFC}

\subsection{Definition and results}

We compute the fermion part of
the SF coupling \cite{Sint:1995ch}
(we set $L=T$ as usual)
at one-loop order $p_{1,1}(L/a)$
for the massless Neuberger-Dirac operator.
The one-loop coefficient
\footnote{Formally we should define the coupling
\be
p_{1,1}(L/a)
=
\left.
\frac{1}{2k}
\frac{\pal}{\pal\eta}
\ln \det[{\Dov^{\dag}\Dov}]
\right|_{\eta=\nu=0},
\ee
in order to properly define the determinant in the continuum limit,
but on the lattice the above form is
equivalent to eq.(\ref{eqn:p11}) due to $\gamma_5$ Hermicity.
}
is given as
\be
p_{1,1}(L/a)
=
\left.
\frac{1}{k}
\frac{\pal}{\pal\eta}
\ln \det{\Dov}
\right|_{\eta=\nu=0},
\label{eqn:p11}
\ee
with a normalization
\be
k=12(L/a)^2[\sin(\gamma)+\sin(2\gamma)],
\mbox{  }
\gamma=\frac{1}{3}\pi(a/L)^2.
\ee
In the actual calculation,
we expand the $\eta$ derivative
and use the fact that the determinant
is factorized to the individual spatial momentum $\pbf$ and color sector $b$,
\bea
p_{1,1}(L/a)
&=&
\frac{1}{k}
{\rm Tr}
\left[
\Dov^{-1}\frac{\pal\Dov}{\pal\eta}
\right]
\non\\
&=&
\frac{1}{k}
\sum_{\pbf}
\sum_{b=1}^{3}
{\rm tr}
\left[
(\Dov^b)^{-1}(\pbf)\frac{\pal\Dov^b(\pbf)}{\pal\eta}
\right].
\label{eqn:p11expand}
\eea
Since $\Dov$ is not the block tri-diagonal in the time
and the spinor index,
unfortunately
we can not use the nice recurrence formula \cite{Sint:1995ch}
which was used in the case of the Wilson-Dirac fermion.
Therefore, we have to evaluate the inverse of $\Dov$ directly by making use of
a solver routine, and multiply with
the $\eta$ derivative of $\Dov$ to take the trace.
The trace in eq.(\ref{eqn:p11expand}),
${\rm tr}$,
concerns with the spinor and the time indices.

To compute the one-loop SF coupling,
we need to take the $\eta$ derivative of $\Dov$.
This can be done analytically.
To this end, we have to evaluate,
\be
\dot{f}_N(X)
=
\sum_{k=0}^{N}\dot{T}_k(Y)c_k,
\label{eqn:recurrencedf}
\ee
where the dotted defines the derivative with respect to $\eta$.
This summation can be evaluated by
another recurrence relation besides the one
needed to compose $\Dov$ itself (the Clenshaw recurrence relation).
The additional recurrence relation can be derived
from the three terms recurrence formula for the Chebyshev polynomials
and its $\eta$ derivative formula.

We compute $p_{1,1}$ on the lattices of size $L/a=4,...,48$.
The results are summarized in Table \ref{tab:p11}
for $s$ parameters, $s=0.0,0.5$
and $\theta=0,\pi/5$.
In order to estimate rounding off errors,
we perform two methods of the approximation
to the inverse square root as mentioned in Section \ref{subsec:build},
the minimax and the Chebyshev interpolation.
In the table, we show nine significant digits where
both approximations agree with each other.
Even though we have used double precision arithmetic
and been demanding $10^{-13}$ precision for the inverse square root,
we lose three to four digits
in the summation step of all momentum and color sectors
in eq.(\ref{eqn:p11expand}).

\subsection{Coefficients of Symanzik's expansion}
From the Symanzik's analysis of the cutoff dependence
of Feynman diagrams on the lattice,
one expects that the one-loop coefficient has
an asymptotic expansion
\begin{equation}
p_{1,1}(L/a)
=
\sum^{\infty}_{n=0}
(a/L)^n[A_n + B_n \ln (L/a)].
\label{eqn:p11symanzik}
\end{equation}
We can reliably extract first several coefficients
by making use of the method in Ref.~\cite{Bode:1999sm}.

For the usual renormalization of the coupling constant,
$B_0$ should be $2b_{0,1}$ where $b_{0,1}$ is
the fermion part of the one-loop coefficient
of $\beta$-function for $N_{\rm f}$ flavors QCD,
\bea
b_0
&=&
b_{0,0}+N_{\rm f}b_{0,1},
\\
b_{0,0}
&=&
\frac{11}{(4\pi)^2},
\\
b_{0,1}
&=&
-\frac{2}{3}
\frac{1}{(4\pi)^2}.
\eea
We confirmed $B_0=2b_{0,1}=-0.00844343...$
to three or four significant digits for all cases ($\theta=0,\pi/5$ and
$-0.6\le s \le 0.6$).
When the tree-level O($a$) improvement is realized,
we expect that $B_1 = 0$ holds.
We check this to $10^{-2}$ or $10^{-4}$ in all cases.
This shows that even though
we have been using the approximate formula
of the boundary coefficient in eq.(\ref{eqn:ctreeL}),
it works well to achieve the tree-level O($a$) improvement
to the precision here.
In the following analysis we set exact values $B_0= -1/(12 \pi^2)$
and $B_1=0$.

\begin{table}[!t]
\begin{center}
$\begin{array}{|r|l|l|l|l|}
     \hline\hline
&\multicolumn{2}{|c|}{\theta=0}
&\multicolumn{2}{|c|}{\theta=\pi/5} \\ \hline
 s & \multicolumn{1}{|c|}{A_0}
   & \multicolumn{1}{|c|}{A_1}
   & \multicolumn{1}{|c|}{A_0}
   & \multicolumn{1}{|c|}{A_1} \\
     \hline
-0.6& 0.016944( 7) & -0.021( 2) &0.015562( 8) &-0.021( 3) \\
    & 0.016937^\ast&            &0.015555^\ast & \\
    \hline
-0.5& 0.015712( 6) & -0.020( 2) &0.014330( 5) &-0.020( 2) \\
    & 0.015708^\ast&            &0.014326^\ast & \\
    \hline
-0.4& 0.014754( 5) & -0.019( 1) &0.013373( 3) &-0.019( 1) \\
    & 0.014751^\ast&            &0.013370^\ast & \\
    \hline
-0.3& 0.013992( 4) & -0.019( 1) &0.012610( 3) &-0.0187( 7)\\
    & 0.013990^\ast&            &0.012609^\ast & \\
    \hline
-0.2& 0.013385( 4) & -0.019( 1) &0.012003( 2) &-0.0187( 6)\\
    & 0.013383^\ast&            &0.012002^\ast & \\
    \hline
-0.1& 0.012912( 4) & -0.019( 1) &0.011530( 2) &-0.0188( 5)\\
    & 0.012911^\ast&            &0.011529^\ast & \\
    \hline
 0.0& 0.012567( 3) & -0.0192( 9)&0.011185( 1) &-0.0191( 4)\\
    & 0.012566^\ast&            &0.011185^\ast & \\
    \hline
 0.1& 0.012354( 3) & -0.0198( 9)&0.010972( 1) &-0.0197( 4)\\
    & 0.012353^\ast&            &0.010972^\ast & \\
    \hline
 0.2& 0.012287( 3) & -0.0207( 9)&0.010905( 2) &-0.0206( 4)\\
    & 0.012285^\ast&            &0.010904^\ast & \\
    \hline
 0.3& 0.012390( 3) & -0.0222( 8)&0.011008( 2) &-0.0220( 4)\\
    & 0.012388^\ast&            &0.011007^\ast & \\
    \hline
 0.4& 0.012704( 3) & -0.024( 1) &0.011322( 2) &-0.0241( 6)\\
    & 0.012703^\ast&            &0.011321^\ast & \\
    \hline
 0.5& 0.013293( 7) & -0.028( 2) &0.011912( 3) &-0.0275( 9)\\
    & 0.013292^\ast&            &0.011911^\ast & \\
    \hline
 0.6& 0.01426( 2)  & -0.032( 7) &0.012880( 3) &-0.033( 1) \\
    & 0.014259^\ast&            &0.012878^\ast & \\
    \hline\hline
  \end{array}$
\end{center}
\caption{The coefficients of asymptotic expansion
for $-0.6\le s\le 0.6$ for $\theta=0,\pi/5$.
The value of $A_0$ with $^\ast$ in the last line in each $s$ parameter block
are the values from the previous calculations
\cite{Sint:1995ch,Alexandrou:1999wr}.
The error for those values should be on the last digit.
}
\label{tab:A0A1}
\end{table}

$A_0$ gives an information about a ratio of $\Lambda$-parameters,
and we show the obtained values in Table \ref{tab:A0A1}.
By combining the previous results from Ref.~\cite{Sint:1995ch,Alexandrou:1999wr},
the values of $A_0$ can be obtained, and
are shown in the second line (numbers with $\ast$) in each $s$
in Table \ref{tab:A0A1} for $\theta=0,\pi/5$.
We observe excellent agreements within errors
for all $s$ and $\theta$ parameters we investigated.

To achieve one-loop O($a$) improvement,
$A_1$ is needed to determine the coefficient
of the fermion part of the boundary
counterterm, $c_{\rm t}^{(1,1)}$ \cite{Sint:1995ch}.
The resulting values are shown in Table \ref{tab:A0A1}.
No $\theta$ dependence on the $A_1$ is observed beyond errors.
The absolute value of $|A_1|=0.02-0.03$ of the Neuberger-Dirac
operator is roughly factor two smaller than
that of Wilson type fermion, $|A_1|=0.038282(2)$ \cite{Sint:1995ch}.
If one imposes an improvement condition
\cite{Sint:1995ch}, one finds that
\be
c_{\rm t}^{(1,1)}=A_1/2.
\ee
For future reference, we provide interpolation formula of
the $c_{\rm t}^{(1,1)}$ as a polynomial of $s$
for $\theta=\pi/5$ where errors are smaller than $\theta=0$ case
\be
c_{\rm t}^{(1,1)}
=
-0.00958-0.00206s-0.00484s^2-0.00748s^3-0.01730s^4,
\label{eqn:ct11}
\ee
for $-0.6 \le s \le 0.6$.

\section{Lattice artifacts of the step scaling function to one-loop order}
\label{sec:SSF}
In this section, we investigate lattice artifacts of
the step scaling function (SSF) \cite{Luscher:1991wu}
$\sigma(2,u)$, which describes the evolution
of the running coupling $\bar{g}^2(L)=u$
under changes of scale $L$ by a factor $2$,
\be
\sigma(2,u)=\bar{g}^2(2L),
\hspace{5mm}
u=\bar{g}^2(L).
\ee
The lattice version of the step scaling function is
denoted by $\Sigma(2,u,a/L)$.

Perturbative estimate of the lattice artifacts
of the step scaling function
can be studied by expanding a relative deviation
\be
\delta (u,a/L) 
\equiv
\frac{\Sigma(2,u,a/L) 
 - \sigma(2,u)}{\sigma(2,u)}
= 
\delta_1(a/L) u  
+ O(u^2).
\ee
The one-loop deviation, $\delta_1(s,a/L)$,
is decomposed into pure gauge and fermion part \cite{Sint:1995ch},
\be
\delta_1(a/L)
=
\delta_{1,0}(a/L)
+N_{\rm f}
\delta_{1,1}(a/L).
\ee
We are currently only interested in the fermion part.
The fermion part of the one-loop deviation 
$\delta_{1,1}(a/L)$ in terms of the
one-loop coefficient of the SF coupling $p_{1,1}$
is given by
\begin{equation}
\delta_{1,1}(a/L)
= 
 p_{1,1}(2L/a)
-p_{1,1}(L/a)
-2 b_{0,1} \ln(2).
 \label{eqn:deviation1}
\end{equation}
Depending on the value of the
boundary counter term $c_{\rm t}^{(1,1)}$,
we denote with $\delta^{(0)}_{1,1}(a/L)$
as the tree level O($a$) improved version
with $c_{\rm t}^{(1,1)}=0$,
and $\delta^{(1)}_{1,1}(a/L)$
the  one-loop O($a$) improved one
for $c_{\rm t}^{(1,1)}=A_1/2$.

\begin{figure}[!t]
 \begin{center}
  \begin{tabular}{cc}
  \resizebox{80mm}{!}{\includegraphics{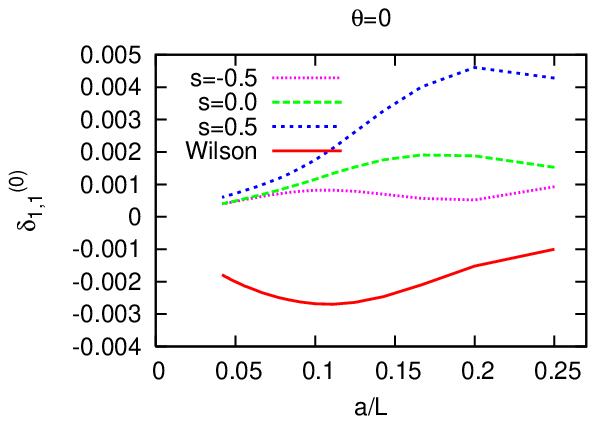}}
  &
  \resizebox{80mm}{!}{\includegraphics{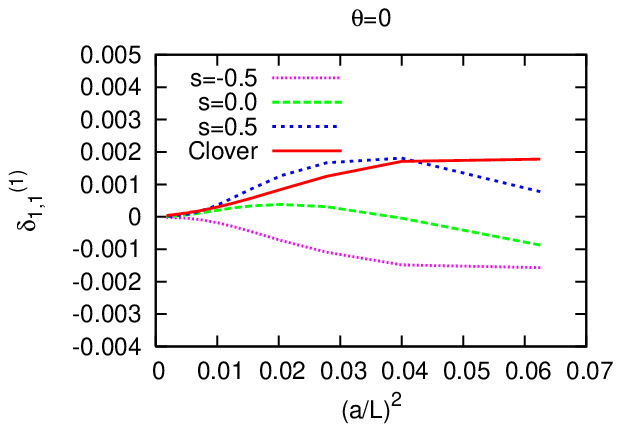}}
  \\
  \resizebox{80mm}{!}{\includegraphics{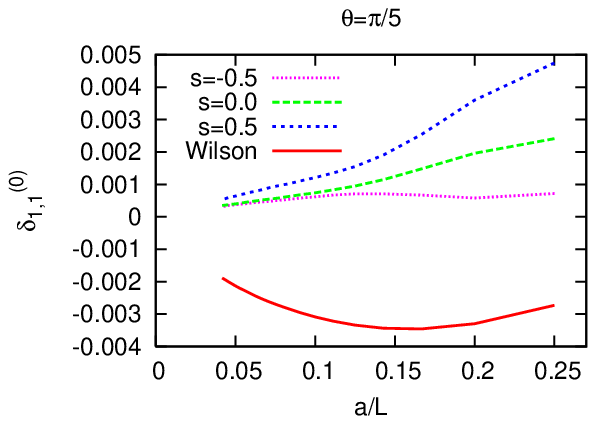}}
  &
  \resizebox{80mm}{!}{\includegraphics{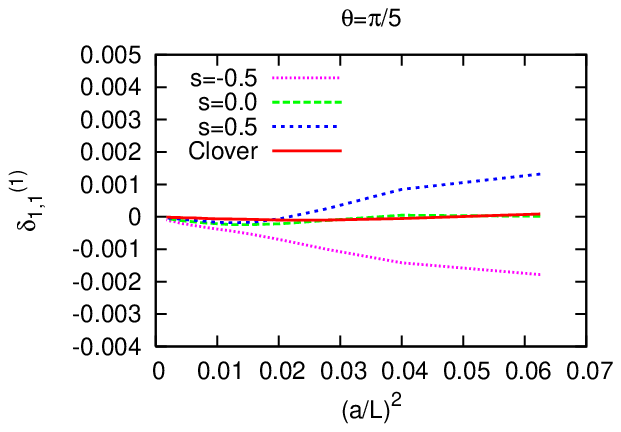}}
  \\
  \end{tabular}
  \caption{We show the relative deviation with the various actions for tree level
$O(a)$ improvement, $\delta_{1,1}^{(0)}$ (Left),
and one-loop $O(a)$ improvement, $\delta_{1,1}^{(1)}$ (Right),
as a function of $a/L$ and $(a/L)^2$ respectively.
Upper part is for $\theta=0$, and lower is for $\theta=\pi/5$.
For comparison, those of the Wilson-Dirac fermion
with $c_{\rm t}^{(1,1)}=0$ and the clover fermion with
$c_{\rm t}^{(1,1)}=0.019141$ \cite{Sint:1995ch}
are included in the plot of $\delta_{1,1}^{(0)}$ and
$\delta_{1,1}^{(1)}$ respectively.
  \label{fig:deviation}}
 \end{center}
\end{figure}

We show numerical results of the one-loop deviation in
Table \ref{tab:deviation} and plots in Figure \ref{fig:deviation},
where we include those of the Wilson-Dirac and the clover action
for comparison \cite{Sint:1995ch}.
In the case of the clover action, $c_{\rm t}^{(1,1)}$ is set
to be the proper value to achieve one-loop $O(a)$ improvement,
and for the Wilson fermion
it is set that $c_{\rm t}^{(1,1)}=0$.
We observe that the lattice artifacts
for the Neuberger-Dirac operator are
comparable to those of the clover action.
As in the case of the clover action,
the Neuberger-Dirac operator
has less lattice artifacts for the case of $\theta=\pi/5$
than $\theta=0$.

\section{Conclusion and outlook}
\label{sec:conclusion}

In this paper, we have explored the free Neuberger-Dirac operator on the SF.
We investigated the spectrum of the operator,
and then we confirmed that the spectrum of $\Dov$ is enclosed by
the unit circle and the spectrum of $\Dov^{\dag}\Dov$
has the expected scaling behavior ($1/L^2$)
and the correct continuum limit.
We also performed the one-loop computation
of the SF coupling by making use of the operator.
We confirmed the universality, and
the fermion part of the O($a$)
boundary counterterm at one-loop order, $c_{\rm t}^{(1,1)}$
is determined.
The formula in eq.(\ref{eqn:ct11}),
$c_{\rm t}^{(1,1)}$ as a function of $s$,
might be useful for future simulations.
By making use of the one-loop results,
we estimated the lattice artifacts of the SSF.
It turns out that the size of the lattice artifacts
for the Neuberger-Dirac operator is
almost the same as that of the clover action.
Thus, we may expect small lattice artifacts
for the non-perturbative SSF of the Neuberger-Dirac operator,
as in the case of the clover action \cite{Dellamorte:2005}.
In Appendix \ref{sec:ctree}, we demonstrate that
the choice of the boundary coefficient $c=1+s$
given in \cite{Luscher:2006df}
at tree level is almost optimal.
This formula may be precise enough for actual simulations.

We exclusively considered the massless case.
Although the massive case can be explored,
we leave it as a future task.
Comparison of scaling behavior
with the other formalism \cite{Taniguchi:2004gf}
and a consistency check is also interesting.
Before starting non-perturbative computations,
we have to compute some improvement coefficients to one or two loop order.
Furthermore perturbative calculations of the renormalization factors
are in a to-do list.
By combining techniques in Ref.~\cite{Takeda:2007dt},
a two-loop calculation of the SF coupling 
including the Neuberger-Dirac operator as a fermion part
may also be feasible.

Apart from the one-loop computations,
next target would be a computation of
the renormalization constant of
the flavor singlet scalar density $Z_S$ non-perturbatively,
since the bare quark condensate in two flavor QCD
was already computed by JLQCD
\cite{Fukaya:2007fb}\footnote{They performed a non-perturbative
renormalization by the RI/MOM scheme.
In this paper here, we are talking
about the non-perturbative renormalization
by making use of SF scheme.}.
Due to the chiral symmetry,
$Z_S$ is identical to the renormalization
constant for the non-singlet flavor pseudo scalar density,
$Z_P=Z_S$ \cite{Sint:1999ke,Hasenfratz:1998jp}.
Actually, the non-perturbative renormalization
group running of $Z_P$ is already known in
Ref.~\cite{DellaMorte:2005kg} for the SF scheme.
A missing piece to obtain the renormalization group invariant
quark condensate is a low energy matching factor,
$Z_P(g_0,\mu=1/L_{\rm max})$ in the SF scheme
for the overlap fermion.
This is an urgent and possibly doable task in the near future.

\section*{Acknowledgments}
We would like to thank Bj\"orn Leder, Stefan Schaefer and Ulli Wolff
for critical reading and giving comments for the manuscript.
I am also grateful to Oliver B\"ar, Michele Della Morte, and Rainer Sommer
for helpful discussions.
This paper is motivated by Discussion Seminar (DS)
which is organized by Ulli Wolff at Humboldt Universit\"at
and DESY Zeuthen.
We thank all participants of the DS for
having discussions and sharing nice atmosphere.
The work is supported in the framework of SFB Transregio 9
of the Deutsche Forschungsgemeinschaft (DFG).
We also thank FLAVIAnet for financial support.

\appendix
\section{Free Wilson-Dirac operator on the SF
in time-momentum representation
with non-zero background field}
\label{sec:Dw}

In the presence of the background gauge field \cite{Luscher:1993gh},
the free part of the Wilson fermion action
in time-momentum space has a form \cite{Sint:1995ch}
\be
S_{\rm w}^0
=
\frac{1}{L^3}
\sum_{\pbf}\sum_{x_0,y_0}^{}
\bar{\psi}(-\pbf,x_0) (\Dw(\pbf;x_0,y_0) + \delta_{x_0,y_0}m_0) \psi(\pbf,y_0),
\ee
with the boundary conditions
\bea
P_+\psi(\pbf,0)&=&0,
\hspace{5mm}
P_-\psi(\pbf,T) = 0,
\\
\bar\psi(\pbf,0)P_-&=&0,
\hspace{5mm}
\bar\psi(\pbf,T)P_+=0.
\eea
The massless part of the
Wilson-Dirac operator on the SF is given as
\be
a \Dw^{bc}(\pbf;x_0,y_0)
=\{
-P_-\delta_{x_0+a,y_0}
+h^b(\pbf;x_0)\delta_{x_0,y_0}
-P_+\delta_{x_0-a,y_0}
\}\delta_{bc},
\ee
where indices $b,c$ refer to color and $h^b(\pbf;x_0)$ is given by
\be
h^b(\pbf;x_0)
=
1
+\frac{1}{2}\sum_{k=1}^{3} \hq_k^b(x_0)^2
+i\sum_{k=1}^{3} \tq_k^b(x_0)\gamma_k,
\ee
with
\bea
\tq_k^b(x_0)
&=&
 \sin(aq_k^b(x_0)  ),
\\
\hq_k^b(x_0)
&=&
2\sin(aq_k^b(x_0)/2),
\eea
and
\bea
q_k^b(x_0)
&=&
w_b x_0+r_k^b,
\\
w_b
&=&
(\phi_b^{\prime}-\phi_b)/L^2,
\\
r_k^b
&=&
p_k + \phi_b /L.
\eea
The spatial component of the momentum ${\bf p}$ is given by
\be
p_k
=
(2\pi n_k + \theta)/L,
\ee
for $n_k=0,\cdots,L/a-1$.
The boundary phases $\phi_b$ and $\phi^{\prime}_b$ are
given by
\begin{align}
\phi_1
&=
\eta - \frac{\pi}{3},
&
\phi^{\prime}_1
&=
-\phi_1 - \frac{4\pi}{3},
\\
\phi_2
&=
\eta(-\frac{1}{2}+\nu),
&
\phi^{\prime}_2
&=
-\phi_3 + \frac{2\pi}{3},
\\
\phi_3
&=
\eta(-\frac{1}{2}-\nu) + \frac{\pi}{3},
&
\phi^{\prime}_3
&=
-\phi_2 + \frac{2\pi}{3}.
\end{align}
Switching off the phases
$\phi_b=\phi^{\prime}_b=0$ correspond to the zero background field.
One can consider the effect of the non-zero background field
a shift for the spatial momentum.
The $\theta$ also affect as a constant
shift for the momentum in a global manner,
on the other hands,
the background field provides a time dependent (local) shift.

In a $4 (T/a-1) \times 4 (T/a-1)$ matrix expression,
the Wilson-Dirac operator is represented by
\bea
\lefteqn{a \Dw^b(\pbf)}
\label{eqn:matrixDw}\\
&=&
\left[
\begin{array}{ccccccc}
h^b(\pbf;a)&  -P_-      &  0         &\cdots&\cdots&  0        &  0        \\
-P_+       &h^b(\pbf;2a)&  -P_-      &0     &\cdots&  \cdots   &  0   \\
0          &  -P_+      &h^b(\pbf;3a)&-P_-  &0     &  \cdots   &  \vdots   \\
\vdots     &  0         &  -P_+      &\ddots&\ddots&  \ddots   &  \vdots   \\
\vdots     &  \ddots    &  \ddots    &\ddots&\ddots&  -P_-     &  0        \\
0          &  \cdots    &  \cdots    &0     &-P_+  &h^b(\pbf;T-2a) & -P_-  \\
0          &  0         &  \cdots    &\cdots&0     & -P_+      &h^b(\pbf;T-a)\\
\end{array}
\right].
\non
\eea

\section{Determination of boundary coefficient $c$ at the tree level}
\label{sec:ctree}

The boundary coefficient $c$ in the kernel
of the overlap operator on the SF is
expanded in terms of the coupling constant $g_0^2$,
\be
c(g_0^2)
=
c^{(0)}
+
c^{(1)}g_0^2
+
O(g_0^4).
\ee
In this appendix, we determine the tree coefficient $c^{(0)}$
which depends on $s$ parameter.
Ref.~\cite{Luscher:2006df} gives
the formula in eq.(\ref{eqn:ctreeL}),
and we will examine it carefully here.

We consider the SF with $T=2L$ and $\theta=0$
in the presence of the non-zero background field.
The massless Neuberger-Dirac operator
is assumed also in this appendix.
The basic correlation functions \cite{Luscher:1996sc} we use are given by
\bea
f_A(x_0)
&=&
-a^6 \sum_{{\bf y},{\bf z}}
\frac{1}{3}
\langle
A_0^a(x) \bar{\zeta}({\bf y})\gamma_5 \frac{1}{2}
\tau^a \zeta({\bf z})
\rangle,
\\
f_P(x_0)
&=&
-a^6 \sum_{{\bf y},{\bf z}}
\frac{1}{3}
\langle
P^a(x) \bar{\zeta}({\bf y})\gamma_5 \frac{1}{2}
\tau^a \zeta({\bf z})
\rangle,
\eea
where boundary fields \cite{Luscher:2006df} are given by
\bea
\zeta({\bf x})
&=&
U(x,0)|_{x_0=0} P_- \psi(x)|_{x_0=a},
\\
\bar{\zeta}({\bf x})
&=&
\bar{\psi}(x)|_{x_0=a} P_+ U(x,0)^{-1}|_{x_0=0}.
\eea

At the tree level, $f_A(x_0)$ $f_P(x_0)$ are given as
\bea
f_{\Gamma}^{(0)}(x_0)
&=&
\sum_{\alpha=1}^3 f_{\Gamma,\alpha}^{(0)}(x_0),
\\
f_{\Gamma,\alpha}^{(0)}(x_0)
&=&
\frac{1}{2}{\rm tr}
\left[
P_+ \Gamma P_- S_\alpha({\bf p=0};a,x_0)
\Gamma S_\alpha({\bf p=0};x_0,a)
\right],
\eea
with $\Gamma=\gamma_0\gamma_5,\gamma_5$ for
$f_A^{(0)}$ and $f_P^{(0)}$
respectively\footnote{One has to use improved operators,
and this is equivalent to replace the propagator
$D^{-1}\rightarrow (1-\bar{a}D/2)D^{-1}$
in the case of massless.
The second term turns out to be a contact term,
and drops when one considers a correlation function
whose insertion points are separated like here.},
and $S_{\alpha}(\pbf;x_0,y_0)$ is a free
propagator in time-momentum space.
The trace in the above equation are
over the Dirac spinor indices only.
The $\alpha$ refers to color.
We chose a ratio $f_A^{(0)}(x_0)/f_P^{(0)}(x_0)$ at a middle point $x_0=T/2$
as an observable to impose the improvement condition.
We compute the quantity $f_A^{(0)}(T/2)/f_P^{(0)}(T/2)$
for lattice size $L/a=8,10,...,64$ and $-0.6\le s \le 0.6$ and some
range of $c^{(0)}$.
We search $c^{(0)}$ around the target point with width
$\Delta c^{(0)} = 0.0005$.

We extract the order $a$ coefficient $A_1$ from
the Symanzik's expansion for the ratio
\be
\frac{f_A^{(0)}(T/2)}{f_P^{(0)}(T/2)}
=
\sum_{n=0}^\infty (a/L)^n A_n.
\label{eqn:ratio}
\ee
We estimate the error of $A_1$ by making use of the method in
\cite{Bode:1999sm}.

\begin{figure}[!h]
 \begin{center}
   \scalebox{2.}{\includegraphics{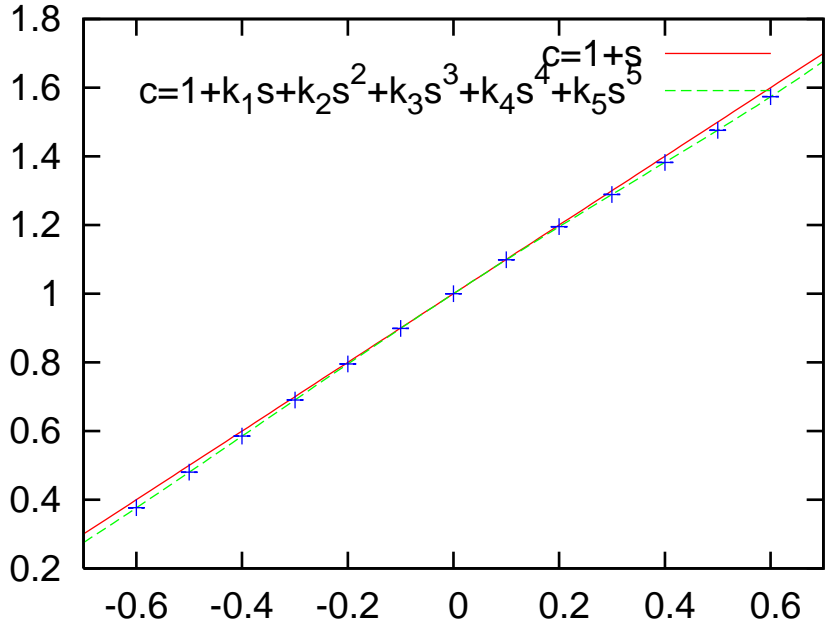}}
 \end{center}
 \caption{The green dashed line represent
$c^{\ast(0)}(s)$ in eq.(\ref{eqn:cast}),
although $\ast$ is not shown in the legend.
The line of $c(s)=1+s$ which is
given in Ref.~\cite{Luscher:2006df} is also shown as solid red.
The error bar of the points are too small to see in this scale.
\label{fig:c-s}}
\end{figure}

We determine $c^{\ast(0)}$ such that $A_1(c^{\ast(0)})=0$ for the range
$-0.6\le s\le 0.6$ (improvement condition).
In Figure \ref{fig:c-s}, we plot 
$c^{\ast(0)}(s)$ as a function of $s$.
By fitting the data points with a functional form
\be
c^{\ast(0)}(s)=1+k_1 s+k_2 s^2+k_3 s^3+k_4 s^4+k_5 s^5,
\label{eqn:cast}
\ee
we obtain
\be
k_1= 1.0002,
k_2=-0.1279,
k_3=-0.0374,
k_4= 0.1616,
k_5= 0.0818.
\label{eqn:ks}
\ee
This curve is also shown in Figure \ref{fig:c-s}.
For larger $s$, a discrepancy between the above formula
and eq.(\ref{eqn:ctreeL}) can be seen, and
their difference is maximally
$10\%$ in the range $-0.6\le s \le 0.6$.
As a consistency check, by making use of the value of
$c$ in eq.(\ref{eqn:cast},\ref{eqn:ks}),
we compute the ratio in eq.(\ref{eqn:ratio})
with $\theta=\pi/5$, and then
$A_1=0$ is confirmed up to $10^{-4}$ in the range
of $s$.

In order to measure an effect of the difference
on a physical quantity,
we compare the one-loop coefficient $p_{1,1}$
with different values of $c$
from the different formulae of $c$ (eq.(\ref{eqn:ctreeL}) and eq.(\ref{eqn:cast}))
at $s=0.5$.
It turns out that a difference in the $p_{1,1}$ is
less than one percent on the lattice size $L/a=4,...,48$.
Furthermore, the resulting
Symanzik's coefficients of $p_{1,1}$
in eq.(\ref{eqn:p11symanzik})
do not change within errors.
Therefore, we concluded that, to the precision in our calculation,
the formula $c=1+s$ is accurate enough
to achieve the tree level O($a$) improvement.

\section{Tables of numerical results}
\label{sec:tables}
\begin{table}[p]
{\small
 \begin{center} 
  \begin{tabular}{|c|rr|rr|rr|c|c|}
  \hline \hline
  \multicolumn{9}{|c|}{$\theta=0$} \\ \hline
  & \multicolumn{2}{c|}{$L/a=6$}
  & \multicolumn{2}{c|}{$L/a=12$}
  & \multicolumn{2}{c|}{$L/a=24$} & & \\
  $n$ & $s= 0.0$ & $s= 0.5$ & $s=0.0$ & $s=0.5$ & $s=0.0$ & $s=0.5$& $b$   & $d$ \\ \hline
  1&       2.707535&       2.544076&       2.353314&       2.293247&       2.228076&       2.202150&  2&  2\\
  2&       5.443956&       5.303784&       5.178180&       5.079915&       5.002836&       4.946708&  2&  2\\
  3&       9.738741&       9.383681&       8.485551&       8.277036&       7.995129&       7.890191&  3&  2\\
  4&      13.327776&      12.783053&      11.160381&      10.827148&      10.347395&      10.185393&  1&  2\\
  5&      13.750702&      13.878968&      12.998595&      12.869161&      12.567207&      12.466990&  3&  2\\
  6&      21.960150&      23.134769&      21.240463&      21.298499&      20.721925&      20.658775&  1&  2\\
  7&      25.715823&      26.315166&      24.501908&      24.335325&      23.730591&      23.572580&  2&  2\\
  8&      26.985530&      27.522077&      25.804837&      25.514692&      24.860285&      24.610036&  2&  2\\
  9&      31.933956&      34.128759&      29.536713&      29.464037&      28.196362&      28.035155&  1&  6\\
 10&      32.231451&      34.580806&      30.746251&      30.712437&      29.585099&      29.393814&  3&  6\\
  \hline \hline
  \multicolumn{9}{|c|}{$\theta=\pi/5$} \\ \hline
  & \multicolumn{2}{c|}{$L/a=6$}
  & \multicolumn{2}{c|}{$L/a=12$}
  & \multicolumn{2}{c|}{$L/a=24$} & & \\
  $n$ & $s= 0.0$ & $s= 0.5$ & $s=0.0$ & $s=0.5$ & $s=0.0$ & $s=0.5$& $b$   & $d$ \\ \hline
  1&       6.011926&       5.711491&       5.218218&       5.083175&       4.924877&       4.861719&  2&  2\\
  2&       6.704204&       6.262644&       5.545431&       5.371364&       5.157911&       5.082355&  1&  2\\
  3&       9.521311&       9.437066&       9.006885&       8.876548&       8.703925&       8.621114&  2&  2\\
  4&      14.212198&      14.477736&      13.828360&      13.748289&      13.497008&      13.413035&  1&  2\\
  5&      17.602694&      17.555801&      15.532119&      15.219759&      14.607734&      14.419874&  3&  2\\
  6&      22.198468&      23.294184&      21.060670&      21.032218&      20.365926&      20.258315&  3&  2\\
  7&      30.072941&      31.462553&      28.145863&      28.028612&      27.065292&      26.897588&  2&  2\\
  8&      31.326562&      32.668102&      29.601109&      29.345781&      28.332464&      28.056341&  2&  2\\
  9&      32.106341&      34.135283&      30.157372&      30.058835&      28.909787&      28.721464&  1&  6\\
 10&      32.344003&      34.955020&      30.320555&      30.299555&      28.987432&      28.807336&  3&  6\\
  \hline \hline
  \end{tabular}
 \end{center}
 \caption{The lowest 10 eigen-values
of the Hermitian operator $L^2 \Dov^{\dag}\Dov$ for $s=0.0,0.5$.
Upper (Lower) panel is for $\theta=0$ ($\theta=\pi/5$).
The $b$ represents the color sector, and the $d$ is for degeneracy for
one flavor.
}
\label{tab:MIN10}
}
\end{table}

\begin{table}[p]
{\footnotesize
 \begin{center} 
  \begin{tabular}{|c|c|c|c|c|}
  \hline \hline
  & \multicolumn{2}{c|}{$\theta=0$} 
  & \multicolumn{2}{c|}{$\theta=\pi/5$}  \\ \hline
  $L/a$ & $s= 0.0$ & $s= 0.5$ & $s=0.0$ & $s=0.5$
  \\ \hline
  4&       -0.0034437717&       -0.0071063261&       -0.0049235088&       -0.0077269840\\
  5&       -0.0050376545&       -0.0080950194&       -0.0059575690&       -0.0077966495\\
  6&       -0.0061863272&       -0.0083445710&       -0.0067717522&       -0.0078856686\\
  7&       -0.0070501647&       -0.0084904737&       -0.0075732720&       -0.0082758128\\
  8&       -0.0077651466&       -0.0086798832&       -0.0083674935&       -0.0088260785\\
  9&       -0.0084065849&       -0.0089725160&       -0.0091311169&       -0.0094410659\\
 10&       -0.0090072060&       -0.0093373962&       -0.0098498710&       -0.0100513225\\
 11&       -0.0095791039&       -0.0097529900&       -0.0105203681&       -0.0106411210\\
 12&       -0.0101258925&       -0.0101900909&       -0.0111448366&       -0.0111969652\\
 13&       -0.0106484681&       -0.0106342058&       -0.0117275060&       -0.0117220888\\
 14&       -0.0111473024&       -0.0110730891&       -0.0122729086&       -0.0122163160\\
 15&       -0.0116231221&       -0.0115016064&       -0.0127852296&       -0.0126837987\\
 16&       -0.0120769723&       -0.0119161854&       -0.0132681309&       -0.0131265368\\
 17&       -0.0125101072&       -0.0123158743&       -0.0137247573&       -0.0135474304\\
 18&       -0.0129238689&       -0.0127003075&       -0.0141578001&       -0.0139483915\\
 19&       -0.0133195990&       -0.0130699396&       -0.0145695681&       -0.0143313843\\
 20&       -0.0136985854&       -0.0134253704&       -0.0149620524&       -0.0146979250\\
 21&       -0.0140620352&       -0.0137673993&       -0.0153369798&       -0.0150494169\\
 22&       -0.0144110635&       -0.0140968164&       -0.0156958565&       -0.0153870369\\
 23&       -0.0147466921&       -0.0144144186&       -0.0160400040&       -0.0157118447\\
 24&       -0.0150698533&       -0.0147209512&       -0.0163705870&       -0.0160247670\\
 25&       -0.0153813962&       -0.0150171154&       -0.0166886380&       -0.0163266371\\
 26&       -0.0156820935&       -0.0153035564&       -0.0169950756&       -0.0166181994\\
 27&       -0.0159726488&       -0.0155808694&       -0.0172907211&       -0.0169001267\\
 28&       -0.0162537034&       -0.0158495996&       -0.0175763116&       -0.0171730277\\
 29&       -0.0165258427&       -0.0161102478&       -0.0178525111&       -0.0174374557\\
 30&       -0.0167896020&       -0.0163632738&       -0.0181199197&       -0.0176939150\\
 31&       -0.0170454713&       -0.0166091003&       -0.0183790818&       -0.0179428671\\
 32&       -0.0172939005&       -0.0168481166&       -0.0186304925&       -0.0181847346\\
 33&       -0.0175353030&       -0.0170806817&       -0.0188746035&       -0.0184199062\\
 34&       -0.0177700595&       -0.0173071278&       -0.0191118283&       -0.0186487395\\
 35&       -0.0179985212&       -0.0175277622&       -0.0193425458&       -0.0188715648\\
 36&       -0.0182210130&       -0.0177428701&       -0.0195671047&       -0.0190886874\\
 37&       -0.0184378353&       -0.0179527167&       -0.0197858262&       -0.0193003902\\
 38&       -0.0186492671&       -0.0181575486&       -0.0199990069&       -0.0195069356\\
 39&       -0.0188555674&       -0.0183575961&       -0.0202069215&       -0.0197085677\\
 40&       -0.0190569772&       -0.0185530739&       -0.0204098244&       -0.0199055137\\
 41&       -0.0192537212&       -0.0187441830&       -0.0206079523&       -0.0200979856\\
 42&       -0.0194460090&       -0.0189311116&       -0.0208015253&       -0.0202861812\\
 43&       -0.0196340368&       -0.0191140363&       -0.0209907486&       -0.0204702859\\
 44&       -0.0198179879&       -0.0192931226&       -0.0211758139&       -0.0206504731\\
 45&       -0.0199980344&       -0.0194685265&       -0.0213569003&       -0.0208269054\\
 46&       -0.0201743375&       -0.0196403945&       -0.0215341758&       -0.0209997357\\
 47&       -0.0203470490&       -0.0198088647&       -0.0217077977&       -0.0211691077\\
 48&       -0.0205163116&       -0.0199740675&       -0.0218779139&       -0.0213351567\\
  \hline \hline
  \end{tabular}
 \end{center}
 \caption{The one loop coefficient $p_{1,1}(L/a)$ for $s=0.0,0.5$ with $\theta=0,\pi/5$.
The last digits may be affected by rounding off errors.}
 \label{tab:p11}
}
\end{table}

\begin{table}[p]
{\footnotesize
 \begin{center} 
  \begin{tabular}{|c|c|c|c|c|c|c|c|c|} 
  \hline \hline
  \multicolumn{9}{|c|}{$\theta=0$} \\ \hline
  & \multicolumn{4}{c|}{$\delta^{(0)}_{1,1}$} 
  & \multicolumn{4}{c|}{$\delta^{(1)}_{1,1}$} 
  \\ \hline
  $L/a$ & $s=-0.5$ & $s=0.0$ & $s=0.5$ & Wilson & $s=-0.5$ & $s=0.0$ & $s=0.5$ & Clover
  \\ \hline
 4&0.00093&0.00153&0.00428&-0.00100&-0.00157&-0.00087&0.00078&0.00178\\
 5&0.00052&0.00188&0.00461&-0.00152&-0.00148&-0.00004&0.00181&0.00171\\
 6&0.00057&0.00191&0.00401&-0.00210&-0.00109&0.00031&0.00167&0.00125\\
 7&0.00070&0.00176&0.00327&-0.00246&-0.00073&0.00038&0.00127&0.00085\\
 8&0.00079&0.00154&0.00262&-0.00264&-0.00046&0.00034&0.00087&0.00058\\
 9&0.00082&0.00134&0.00212&-0.00270&-0.00029&0.00027&0.00057&0.00041\\
10&0.00082&0.00116&0.00176&-0.00268&-0.00018&0.00020&0.00036&0.00030\\
11&0.00079&0.00102&0.00151&-0.00263&-0.00012&0.00015&0.00024&0.00023\\
12&0.00075&0.00091&0.00132&-0.00256&-0.00008&0.00011&0.00016&0.00018\\
13&0.00071&0.00082&0.00118&-0.00249&-0.00006&0.00008&0.00011&0.00015\\
14&0.00067&0.00075&0.00108&-0.00241&-0.00004&0.00006&0.00008&0.00012\\
15&0.00063&0.00069&0.00099&-0.00234&-0.00004&0.00005&0.00006&0.00011\\
16&0.00059&0.00064&0.00092&-0.00226&-0.00003&0.00004&0.00005&0.00009\\
17&0.00056&0.00059&0.00086&-0.00219&-0.00003&0.00003&0.00004&0.00008\\
18&0.00053&0.00056&0.00081&-0.00213&-0.00003&0.00002&0.00003&0.00007\\
19&0.00050&0.00052&0.00076&-0.00206&-0.00003&0.00002&0.00003&0.00006\\
20&0.00048&0.00049&0.00072&-0.00200&-0.00002&0.00001&0.00002&0.00006\\
21&0.00045&0.00047&0.00069&-0.00195&-0.00002&0.00001&0.00002&0.00005\\
22&0.00043&0.00045&0.00066&-0.00189&-0.00002&0.00001&0.00002&0.00005\\
23&0.00041&0.00042&0.00063&-0.00184&-0.00002&0.00001&0.00002&0.00004\\
24&0.00040&0.00041&0.00060&-0.00179&-0.00002&0.00001&0.00002&0.00004\\
  \hline \hline
  \multicolumn{9}{|c|}{$\theta=\pi/5$}  \\ \hline
  & \multicolumn{4}{c|}{$\delta^{(0)}_{1,1}$} 
  & \multicolumn{4}{c|}{$\delta^{(1)}_{1,1}$} 
  \\ \hline
  $L/a$ & $s=-0.5$ & $s=0.0$ & $s=0.5$ & Wilson & $s=-0.5$ & $s=0.0$ & $s=0.5$ & Clover
  \\ \hline
 4&0.00072&0.00241&0.00475&-0.00273&-0.00178& 0.00002& 0.00132& 0.00009\\
 5&0.00058&0.00196&0.00360&-0.00330&-0.00142& 0.00005& 0.00085&-0.00005\\
 6&0.00067&0.00148&0.00254&-0.00346&-0.00100&-0.00011& 0.00025&-0.00010\\
 7&0.00071&0.00115&0.00191&-0.00344&-0.00071&-0.00021&-0.00005&-0.00010\\
 8&0.00071&0.00095&0.00155&-0.00334&-0.00054&-0.00024&-0.00017&-0.00008\\
 9&0.00067&0.00083&0.00135&-0.00322&-0.00044&-0.00024&-0.00018&-0.00007\\
10&0.00062&0.00074&0.00121&-0.00309&-0.00038&-0.00021&-0.00017&-0.00006\\
11&0.00058&0.00068&0.00111&-0.00296&-0.00033&-0.00019&-0.00014&-0.00005\\
12&0.00054&0.00063&0.00102&-0.00284&-0.00029&-0.00017&-0.00012&-0.00004\\
13&0.00051&0.00058&0.00096&-0.00273&-0.00026&-0.00015&-0.00010&-0.00003\\
14&0.00048&0.00055&0.00090&-0.00263&-0.00023&-0.00013&-0.00009&-0.00003\\
15&0.00046&0.00052&0.00084&-0.00253&-0.00021&-0.00012&-0.00007&-0.00003\\
16&0.00044&0.00049&0.00079&-0.00244&-0.00019&-0.00011&-0.00007&-0.00002\\
17&0.00042&0.00047&0.00075&-0.00235&-0.00017&-0.00010&-0.00006&-0.00002\\
18&0.00040&0.00044&0.00071&-0.00227&-0.00015&-0.00009&-0.00005&-0.00002\\
19&0.00039&0.00042&0.00068&-0.00220&-0.00014&-0.00008&-0.00005&-0.00002\\
20&0.00037&0.00040&0.00064&-0.00213&-0.00013&-0.00007&-0.00004&-0.00002\\
21&0.00036&0.00039&0.00062&-0.00206&-0.00012&-0.00007&-0.00004&-0.00001\\
22&0.00035&0.00037&0.00059&-0.00200&-0.00011&-0.00006&-0.00004&-0.00001\\
23&0.00033&0.00036&0.00056&-0.00194&-0.00010&-0.00006&-0.00003&-0.00001\\
24&0.00032&0.00035&0.00054&-0.00189&-0.00009&-0.00005&-0.00003&-0.00001\\
  \hline \hline
  \end{tabular}
 \end{center}
 \caption{The relative deviation $\delta_{1,1}^{(0,1)}$
for $s=-0.5,0.0,0.5$ for $\theta=0$(upper) and $\theta=\pi/5$(lower).
Results for the Wilson and the clover action \cite{Sint:1995ch} are
also included for comparison.
}
 \label{tab:deviation}
}
\end{table}

\newpage
\providecommand{\href}[2]{#2}\begingroup\raggedright\endgroup
\end{document}